\documentclass[11pt, final]{article}

\usepackage{setspace}
\usepackage[all]{nowidow}
\usepackage[english]{babel}
\usepackage{microtype}

\usepackage{amsmath}
\usepackage{amssymb}
\usepackage{bm}
\usepackage{physics}
\usepackage{mathtools}

\usepackage{cite}
\usepackage{authblk}

\usepackage{tikz}
\usepackage{tabularx}
\usepackage{array}
\usepackage{multicol}
\usepackage[font=footnotesize]{caption}
\newcolumntype{C}{>{$}c<{$}}                
\usetikzlibrary{arrows, positioning}

\usepackage{changes}
\usepackage{ifdraft}
\ifdraft{\usepackage[left=5mm, right=45mm, marginparwidth=40mm]{geometry}}{\usepackage{geometry}}

\newcommand{\mb}{\mathbb}
\newcommand{\mc}{\mathcal}
\newcommand{\SL}{\operatorname{\textsl{SL}}}    
\newcommand{\PSL}{\operatorname{\textsl{PSL}}}  
\newcommand{\vn}{V^{\natural}}                  
\newcommand{\vbn}{V \mb B^\natural}             
\newcommand{\hecke}{\mathsf{T}}                 
\newcommand{\diag}{\operatorname{\text{diag}}}
\newcommand{\transpose}{\intercal}              
\numberwithin{equation}{section}

\begin{document}

\title{Modular invariance groups and defect McKay-Thompson series}
\author{Harry Fosbinder-Elkins\thanks{harryfe@uchicago.edu} }
\author{Jeffrey A. Harvey\thanks{j-harvey@uchicago.edu}}
\affil{Kadanoff Center for Theoretical Physics, Enrico Fermi Institute, and Department of Physics \\ University of Chicago \\ Chicago, IL 60637, U.S.A.}

\maketitle

\abstract{It has been known since 1992 that the McKay-Thompson series $T_g(q)$ of the Moonshine module form Hauptmoduln for genus zero subgroups of $SL(2, \mathbb{R})$ \cite{conway_monstrous_1979,thompson_numerology_1979,frenkel_vertex_1989,borcherds_monstrous_1992}. In \cite{lin_duality_2021}, Lin and Shao constructed a series analogous to the McKay-Thompson series (a twined partition function of the Monster CFT), but using a non-invertible topological defect rather than an element of the Monster group $\mathcal{M}$. This ``defect McKay-Thompson series'' was found to be invariant under a genus zero subgroup of $SL(2, \mathbb{R})$, but was shown not to be the Hauptmodul of the subgroup. Nevertheless, one might wonder if a weaker version of Borcherds' theorem holds for non-invertible defects: perhaps defect McKay-Thompson series enjoy genus zero invariance groups in $SL(2, \mathbb{R})$, whether or not they are Hauptmoduln for those groups. Using the decompositions of the monster stress tensor found in \cite{bae_conformal_2021}, we construct several new defect McKay-Thompson series, study their modular properties, and determine their invariance groups in $SL(2, \mathbb{R})$. We discover that many of the invariance groups are not genus zero.}
\clearpage

\section{Introduction}
The phenomenon of monstrous moonshine, conjectured by Conway and Norton in \cite{conway_monstrous_1979}, connects the representation theory of the Monster group $\mc M$ with the theory of modular functions. Specifically, Conway and Norton conjectured that the McKay-Thompson series $T_g(q)$ associated with elements $g$ of $\mc M$ are Hauptmoduln for genus zero subgroups of $\SL(2,\mathbb{R})$. Frenkel, Lepowsky, and Meurman's construction in \cite{frenkel_vertex_1989} of a vertex operator algebra (VOA) with outer automorphism group $\mc M$ established a connection between moonshine and conformal field theory. The genus
zero property was then proved by Borcherds \cite{borcherds_monstrous_1992}. This remarkable correspondence has since led to deep insights into both number theory and the theory of VOAs.

In \cite{lin_duality_2021}, Lin and Shao extended these ideas by introducing the ``defect McKay-Thompson series.'' Instead of computing the partition function of the monster CFT twined by a group element $g$, they considered the insertion of (generically non-invertible) topological defect lines (TDLs) into the partition function. Lin and Shao found an example of a non-invertible TDL in the monster CFT (the Kramers-Wannier defect), and showed that the associated defect McKay-Thompson series enjoyed a genus zero invariance group in $\SL(2, \mb R)$, though it was not the Hauptmodul. This work left open the question of whether all defect McKay-Thompson series share the genus-zero property of their classical counterparts or whether the genus-zero property found in \cite{lin_duality_2021} was a coincidence due to the invariance of the modular functions under $\SL(2, \mb Z)$ subgroups of low level. 

If $\Gamma$ is a discrete subgroup of $ \SL(2, \mb R)$ that is commensurable with $ \SL(2, \mb Z)$, then the quotient of the upper half plane $\mb H$ (plus cusps) by $\Gamma$ is a compact Riemann surface. If the surface has genus zero, then we say that $\Gamma$ is a genus zero subgroup of $\SL(2, \mb R)$. The genus zero property of the McKay-Thompson series of monstrous moonshine is one of the most striking and mysterious aspects of monstrous moonshine. Genus zero groups also play a central role in organizing other instances of moonshine including umbral and penumbral moonshine \cite{Cheng:2012tq,Duncan:2021ith}. Thus, it is of great interest to determine whether or not there is an extension of the genus zero property to defect McKay-Thompson series. 

In this work, we construct a number of new defect McKay-Thompson series, using decompositions of the stress tensor of the monster CFT into those of smaller RCFTs. These decompositions, inspired by Norton's monstralizer pairs, the work of H{\"o}hn \cite{hohn_selbstduale_1995} and subsequent generalizations which are explicated and generalized in \cite{bae_conformal_2021}, allow us to (non-exhaustively) catalog TDLs of the monster theory, as well as their eigenvalues, from which we may construct the defect McKay-Thompson series. We may then use the known modular properties of the smaller VOAs appearing in our decompositions to infer the modular properties of the defect McKay-Thompson series. Doing so, we find that the genus-zero property of the classical McKay-Thompson series does not hold in general for their non-invertible counterparts.

The paper is organized as follows: In section 2, we describe the methodology used to construct and analyze the defect McKay-Thompson series, including the use of monstralizer pairs. Section 3 describes our method for computing the modular invariance groups of the resulting objects, first in $\SL(2, \mb Z)$ and then considering enlargements in $\SL(2, \mb R)$. Finally, section 4 presents our findings on the invariance groups of several defect McKay-Thompson series, detailing both the genus zero and non-genus zero cases, and discusses potential extensions to this work.

\section{TDLs from monstralizers}\label{ss:tdls}
The term ``monstralizer pair'' is due to Norton \cite{norton_anatomy_1998}, who used it to describe subgroups of $\mc M$ which mutually centralize one another. In \cite{bae_conformal_2021}, Bae et al. used Norton's monstralizer pairs as a clue to construct decompositions of the stress tensor of the monster CFT in terms of commuting conformal vectors for chiral subalgebras of $\vn$ whose symmetry groups are those subgroups of $\mc M$ appearing in the monstralizer pair. This definition may be thought of as the ``VOA uplift'' of the group-theoretic definition of monstralizers. More precisely: say the monster module $\vn$ has subVOA (chiral subalgebra) $\mc A$ with commutant $\mc B$, the outer automorphism groups of which form a monstralizing pair of subgroups in the sense of Norton. This implies that the partition function of $\vn$ may be decomposed as
\begin{equation}\label{eq:tdls:j}
    J(\tau) = \sum_{i j} N_{i j} \chi_i^{(\mc A)}(\tau) \chi_j^{(\mc B)}(\tau)
\end{equation}
for some (possibly not all) characters $\chi_i^{(\mc A)}$ of $\mc A$ and $\chi_i^{(\mc B)}$ of $\mc B$; moreover, the conformal vectors $t_{\mc A}(z)$ and $t_{\mc B}(z)$ associated to these chiral subalgebras must combine to form the $c = 24$ stress tensor of the monster CFT:
\begin{equation}
    T(z) = t_{\mc A}(z) + t_{\mc B}(z).
\end{equation}
We then say that $\mc A$ and $\mc B$ form a monstralizer pair at the level of VOAs. Next, we seek out a (modular invariant) RCFT, which we call the monstralizing RCFT, whose chiral algebra is $\mc A$. By modular invariance of Klein's $j$ invariant, the antichiral algebra $\bar{\mc A}'$ of the RCFT must share modular data with $\mc B$:
\begin{equation} \label{eq:tdls:sb}
    S^{(\mc B)} = S^{(\bar{\mc A}')} = S^{(\mc A) *} \qq{and} T^{(\mc B)} = T^{(\bar{\mc A}')} = T^{(\mc A) *},
\end{equation}
and we demand that, upon matching $\mc A \otimes \mc B$ primaries in $\vn$ with $\mc A \otimes \bar{\mc A}'$ primaries in the monstralizing RCFT, the multiplicities $N_{i j}$ are the same. A variety of monstralizer pairs of VOAs are cataloged in \cite{bae_conformal_2021}. In each of those examples the monstralizing RCFT is a finite sum of minimal models.

Given a monstralizer pair of VOAs, suppose that the monstralizing RCFT contains a topological defect line $\mc L^{(\mc A)}$ (preserving $\mc A \otimes \bar{\mc A}'$), whose action on the primaries is
\begin{equation}\label{eq:tdls:la}
    \mc L^{(\mc A)} \phi_{i j}^{(A)} = L_{i j}^{(\mc A)} \phi_{i j}^{(\mc A)}
\end{equation}
where $(i, j)$ are the left- and right-moving indices from (\ref{eq:tdls:j}). More generally, topological defects can act non-diagonally on the primaries, but this type of TDL will not appear in our analysis. Then, using the map between $\mc A \otimes \mc B$ primaries of $\vn$ and $\mc A \otimes \bar{\mc A}'$ primaries of the monstralizing RCFT (call it $\pi$), we may construct a TDL $\mc L$ of the monster CFT by imposing that the action of $\mc L$ on a given $\mc A \otimes \mc B$ primary of $\vn$ (say $\phi_i$) is
\begin{equation}\label{eq:tdls:l}
    \mc L \phi_{i j} = L_{\pi(\phi_{i j})}^{(\mc A)} \phi_{i j}.
\end{equation}
Such a line is topological by construction (acting identically on primaries and descendants), and (\ref{eq:tdls:j}) and (\ref{eq:tdls:sb}) guarantee that it satisfies the Cardy condition since $\mc L^{(\mc A)}$ does so; it therefore forms a valid TDL in the monster CFT. Moreover, $\mc L$ will preserve the subalgebra $\mc A \otimes \mc B$ of $\vn$. In fact, we need not start with a line preserving the maximal left-moving chiral algebra of the monstralizing RCFT. More generally, applying (\ref{eq:tdls:l}) to a defect preserving $\mc A' \otimes \bar{\mc A}'$ in the monstralizing RCFT (with Virasoro $< \mc A' < \mc A$) yields a TDL of the monster CFT preserving $\mc A' \otimes \mc B$. We emphasize, though, that while $\mc L$ and $\mc L^{(\mc A)}$ have the same eigenvalues, they may differ in other respects -- for example, $\mc L$ is generically not a Verlinde line, whether or not $\mc L^{(\mc A)}$ is. At the level of characters, this construction allows us to compute the partition function of the monster CFT twined by the line $\mc L$:
\begin{equation}\label{eq:tdls:zl}
    Z^{\mc L}(\tau) = \sum_i L_{i j}^{(\mc A)} N_{i j} \chi_i^{(\mc A)}(\tau) \chi_j^{(\mc B)}(\tau) \equiv \vec{\chi}^{(\mc A)}(\tau)^\intercal D_{\mc L} \vec{\chi}^{(\mc B)}(\tau)
\end{equation}
where $\vec{\chi}^{(\mc A)}(\tau)$ (resp. $\vec{\chi}^{(\mc B)}(\tau)$) is a vector formed of those characters of $\mc A$ (resp. $\mc B$) which appear in the monstralizer decomposition (\ref*{eq:tdls:j}). The objects $Z^{\mc L}(\tau)$ are known as ``defect McKay-Thompson series,'' a term defined in \cite{lin_duality_2021}. In that paper, Lin and Shao identified two TDLs of the monster theory, one corresponding to the $\mb Z_{2 \text A}$ conjugacy class of the monster and one non-invertible line, analogous to the Kramers-Wannier defect of the Ising CFT. Using the monstralizer pair $\qty(\mb Z_{2 \text A}, 2.\mb B)$ in the notation of \cite{bae_conformal_2021}, we recover the same lines and the associated defect McKay-Thompson series (given in detail in section \ref*{ss:tdls:ising}). We also consider the monstralizer pairs $\qty(\mb Z_{3 \text A}, 3.Fi'_{24})$, $\qty(\mb Z_{4 \text A}, 4.2^{22}.Co_3)$, and $\qty(D_{3 \text A}, Fi_{23})$. In each case we discover new TDLs of the monster theory, and in two cases we are able to show that we have found all TDLs arising from the monstralizer pair under consideration.

\subsection{\boldmath $\qty(\mb Z_{2 \text A}, 2.\mb B)$: Ising CFT} \label{ss:tdls:ising}
The monstralizer pair $(\mb Z_{2 \text A}, 2.\mb B)$ has been known at the level of VOAs since \cite{hohn_selbstduale_1995}, though in different language. In that work, H\"ohn constructed a subVOA of $\vn$ whose automorphism group is the baby monster $\mb B$. Lin and Shao provided a physical explanation in \cite{lin_duality_2021}: the fermionization of the monster CFT under an element of the $\mb Z_{2 \text A}$ conjugacy class is the direct product of H\"ohn's baby monster VOA with a single Majorana-Weyl fermion. Conversely, under bosonization, they showed that the $\mb Z_2$ symmetry of the direct product theory corresponding to fermion parity becomes the non-invertible element of a fusion subcategory of the monster which is isomorphic to the Ising category. In our language, this subcategory arises from the monstralizer pair $\qty(\mb Z_{2 \text A}, 2.\mb B)$, which uplift to VOAs as Virasoro at $c = \frac{1}{2}$ and the baby monster VOA; we may take the associated monstralizing RCFT to be the Ising CFT. The Verlinde lines of the Ising CFT exhaust the fusion category of TDLs - i.e. no other Virasoro-preserving lines exist. They are reproduced (along with their action on the primaries) for convenience in table \ref{tb:tdls:ising}. Note that in this table, as well as those to follow, the top row does not label individual Virasoro primaries of the monster; rather it should be understood to label sets of primaries of $\vn$ with respect to the ``Ising'' chiral subalgebra.

\begin{table}
    \centering
    \begin{tabular}{| C | C C C |}
        \hline
                & 1         & \sigma    & \epsilon  \\ \hline
        1       & 1         & 1         & 1         \\
        \eta    & 1         & 1         & -1        \\
        \mc D   & \sqrt{2}  & -\sqrt{2} & 0         \\ \hline
    \end{tabular}
    \caption{\label{tb:tdls:ising} TDLs of the Ising subcategory within the monster CFT. We emphasize that the labels $1$, $\sigma$, $\epsilon$, which refer to Virasoro primaries in the constext of the Ising CFT, here refer to primaries of $\vn$ under its $c = \frac{1}{2}$ Virasoro chiral subalgebra.}
\end{table}

The $J$-invariant may be written in terms of Ising and baby monster characters as
\begin{equation}
    J(\tau) = \chi_1(\tau) \chi^{(\vbn)}_1(\tau) + \chi_\sigma(\tau) \chi^{(\vbn)}_\sigma(\tau) + \chi_\epsilon(\tau) \chi^{(\vbn)}_\epsilon(\tau)
\end{equation}
where $\qty{1, \sigma, \epsilon}$ refer to the Ising primaries with $h = \qty{0, \frac{1}{16}, \frac{1}{2}}$ respectively, and the baby monster characters $\chi^{(\vbn)}_i$ are related to the Ising characters by the Hecke operator for vector-valued modular forms $\hecke_{47}$ \cite{bae_conformal_2021,harvey_hecke_2018}. To obtain the defect McKay-Thompson series, we make use of equation \ref*{eq:tdls:zl}; we find $q$-expansions for the series $Z^\eta$ and $Z^{\mc D}$ which agree with those in \cite{lin_duality_2021}. The line $\eta$ simply implements the $\mb Z_{2 \text A}$ global symmetry of the monster, so $Z^{\eta}$ is a ``typical'' McKay-Thompson series, of the type studied by Borcherds, and satisfies the genus zero property. On the other hand, the line $\mc D$ is non-invertible, so $Z^{\mc D}$ is not of this type, and is not guaranteed to possess a genus zero invariance group.

\subsection{\boldmath $\qty(\mb Z_{3 \text A}, 3.Fi'_{24})$: $\mb Z_3$ parafermion} \label{ss:tdls:potts}
The next monstralizer pair we consider is $(\mb Z_{3 \text A}, 3.Fi'_{24})$, which uplift to VOAs as the chiral algebra $W_3$ and its commutant in $\vn$, with monstralizing RCFT the $\mb Z_3$ parafermion CFT. Unlike the $\mb Z_{2 A}$ case, the commutant VOA here is not a familiar one (though it was studied in \cite{hohn_mckays_2012}); we shall only need its characters, which are images of the parafermion characters under the action of the Hecke operator $\hecke_{29}$. Also unlike the $\mb Z_{2 A}$ case, the Verlinde lines of the $\mb Z_3$ parafermion CFT do not exhaust its fusion category of topological defects: there are additional lines which commute with the Virasoro algebra at $c = \frac{4}{5}$, but not the extended chiral algebra \cite{haghighat_topological_2023, chang_topological_2019}. Per the discussion in the introduction to this section, to obtain as many monstrous TDLs as possible we should start with TDLs of the $\mb Z_3$ parafermion CFT preserving $(\text{Virasoro} \otimes \overline{W}_3)$. The TDLs in this category and their eigenvalues are given in table \ref*{tb:tdls:potts}.
\begin{table}
    \noindent\makebox[\linewidth][c]{
    \begin{tabular}{| C | C C C C C C C C C C C C |}
        \hline
                 & 1                & Y                 & \Omega            & \tilde{\Omega}   & \epsilon              & X                      & \Phi                   & \tilde{\Phi}          & Z                & Z^*              & \sigma                 & \sigma^*               \\ \hline
        1        & 1                & 1                 & 1                 & 1                & 1                     & 1                      & 1                      & 1                     & 1                & 1                & 1                      & 1                      \\
        \eta     & 1                & 1                 & 1                 & 1                & 1                     & 1                      & 1                      & 1                     & \alpha           & \alpha^2         & \alpha                 & \alpha^2               \\
        \eta^2   & 1                & 1                 & 1                 & 1                & 1                     & 1                      & 1                      & 1                     & \alpha^2         & \alpha           & \alpha^2               & \alpha                 \\
        W        & \varphi          & \varphi           & \varphi           & \varphi          & -\varphi^{-1}         & -\varphi^{-1}          & -\varphi^{-1}          & -\varphi^{-1}         & \varphi          & \varphi          & -\varphi^{-1}          & -\varphi^{-1}          \\
        \eta W   & \varphi          & \varphi           & \varphi           & \varphi          & -\varphi^{-1}         & -\varphi^{-1}          & -\varphi^{-1}          & -\varphi^{-1}         & \alpha \varphi   & \alpha^2 \varphi & -\alpha \varphi^{-1}   & -\alpha^2 \varphi^{-1} \\
        \eta^2 W & \varphi          & \varphi           & \varphi           & \varphi          & -\varphi^{-1}         & -\varphi^{-1}          & -\varphi^{-1}          & -\varphi^{-1}         & \alpha^2 \varphi & \alpha \varphi   & -\alpha^2 \varphi^{-1} & -\alpha \varphi^{-1}   \\
        N        & \sqrt{3}         & -\sqrt{3}         & -\sqrt{3}         & \sqrt{3}         & -\sqrt{3}             & \sqrt{3}               & \sqrt{3}               & -\sqrt{3}             & 0                & 0                & 0                      & 0                      \\
        W N      & \sqrt{3} \varphi & -\sqrt{3} \varphi & -\sqrt{3} \varphi & \sqrt{3} \varphi & \sqrt{3} \varphi^{-1} & -\sqrt{3} \varphi^{-1} & -\sqrt{3} \varphi^{-1} & \sqrt{3} \varphi^{-1} & 0                & 0                & 0                      & 0                      \\ \hline
    \end{tabular}}
    \caption{\label{tb:tdls:potts} The fusion subcategory of the monster CFT inherited from the $\mb Z_3$ parafermion. Here $\alpha = (-1)^{2/3}$, and $\varphi = \frac{1 + \sqrt{5}}{2}$. The primaries are labeled by the corresponding Virasoro primaries of the $\mb Z_3$ parafermion CFT, in agreement with the labels in \cite{chang_topological_2019,haghighat_topological_2023}.}
\end{table}
The lines labeled $W$, $\eta W$, $\eta^2 W$, $N$, and $W N$ are non-invertible. We pause to note that, since the modular invariant of the theory is not diagonal (with respect to Virasoro), the matrices $D_{\mc L}$ of (\ref{eq:tdls:zl}) representing the partition function of the theory twined by the various TDLs will also not be diagonal. In particular, given a $\mb Z_3$ parafermion TDL $\mc L$ whose eigenvalues on the Virasoro primaries are denoted $L^{\mc L}_\phi$, the partition function for the monster CFT twined by the associated TDL is, using (\ref{eq:tdls:zl}),
\begin{equation}
    Z^{\mc L}(\tau) = \vec{\chi}^{(W_3)}(\tau)^\intercal \smqty(
        L^{\mc L}_I                & L^{\mc L}_{\Omega} &                          &                  &                                 &                                           &   &   &   &   \\
        L^{\mc L}_{\tilde{\Omega}} & L^{\mc L}_{Y}      &                          &                  &                                 &                                           &   &   &   &   \\
                                   &                    & L^{\mc L}_{\epsilon}     & L^{\mc L}_{\Phi} &                                 &                                           &   &   &   &   \\
                                   &                    & L^{\mc L}_{\tilde{\Phi}} & L^{\mc L}_{X}    &                                 &                                           &   &   &   &   \\
                                   &                    &                          &                  & L^{\mc L}_{Z} + L^{\mc L}_{Z^*} &                                           &   &   &   &   \\
                                   &                    &                          &                  &                                 & L^{\mc L}_{\sigma} + L^{\mc L}_{\sigma^*} &   &   &   &   \\
                                   &                    &                          &                  &                                 &                                           & 0 &   &   &   \\
                                   &                    &                          &                  &                                 &                                           &   & 0 &   &   \\
                                   &                    &                          &                  &                                 &                                           &   &   & 0 &   \\
                                   &                    &                          &                  &                                 &                                           &   &   &   & 0
    ) \vec{\chi}^{(\tilde{W}_3)}(\tau)
\end{equation}
where $\vec{\chi}^{(W_3)}(\tau)$ are the $\mb Z_3$ parafermion characters and $\vec{\chi}^{(\tilde{W}_3)}$ are those of the commutant VOA (cf. equation 4.20 of \cite{haghighat_topological_2023}). Once again, using the Hecke operator for vector-valued modular forms, we are able to compute $q$-expansions for the defect McKay-Thompson series associated to each line. $Z^{\mc \eta}$ and $Z^{\mc \eta^2}$ are the usual McKay-Thompson series for the $\mb Z_{3 A}$ conjugacy class; the other series are novel.

\subsection{\boldmath $\qty(\mb Z_{4 \text A}, 4.2^{22}.Co_3)$: $\mb Z_4$ parafermion}
Our next pair of monstralizer VOAs are the chiral algebra of the $\mb Z_4$ parafermion CFT and its commutant in $\vn$. Similar to the $\mb Z_{3 A}$ case, we may take the monstralizing RCFT to be the $\mb Z_4$ parafermion CFT. As in the $\mb Z_3$ case, the $\mb Z_4$ parafermion CFT contains topological defects which commute with Virasoro but not the fully extended chiral algebra (in fact, infinitely many). However, to find the defect McKay-Thompson series associated to these, our method would require us to decompose the modular invariant of the $\mb Z_4$ parafermion CFT into finitely many Virasoro characters. As the theory is irrational under Virasoro, this is not possible, so we satisfy ourselves with TDLs preserving the fully extended chiral algebra (under which the theory is rational). These lines are given in table \ref*{tb:tdls:z4} \cite{haghighat_topological_2023}.
\begin{table}[]
    \centering
    \begin{tabular}{| C | C C C C C C C C C C |}
        \hline
        (l, m)      & (4, 4)    & (4, 2)        & (4, 0)    & (4, -2)       & (3, 3)    & (3, 1)    & (3, -1)   & (1, 1)    & (2, 0)    & (2, 2)    \\ \hline
        1           & 1         & 1             & 1         & 1             & 1         & 1         & 1         & 1         & 1         & 1         \\
        \theta      & 1         & -1            & 1         & -1            & i         & -i        & i         & -i        & 1         & -1        \\
        \theta^2    & 1         & 1             & 1         & 1             & -1        & -1        & -1        & -1        & 1         & 1         \\
        \theta^3    & 1         & -1            & 1         & -1            & -i        & i         & -i        & i         & 1         & -1        \\
        X           & \sqrt{3}  & \sqrt{3} i    & -\sqrt{3} & -\sqrt{3} i   & \beta     & \beta^3   & -\beta    & -\beta^3  & 0         & 0         \\
        \theta X    & \sqrt{3}  & -\sqrt{3} i   & -\sqrt{3} & \sqrt{3} i    & \beta^3   & \beta     & -\beta^3  & -\beta    & 0         & 0         \\
        \theta^2 X  & \sqrt{3}  & \sqrt{3} i    & -\sqrt{3} & -\sqrt{3} i   & -\beta    & -\beta^3  & \beta     & \beta^3   & 0         & 0         \\
        \theta^3 X  & \sqrt{3}  & -\sqrt{3} i   & -\sqrt{3} & \sqrt{3} i    & -\beta^3  & -\beta    & \beta^3   & \beta     & 0         & 0         \\
        Y           & 2         & 2             & 2         & 2             & 0         & 0         & 0         & 0         & -1        & -1        \\
        \theta Y    & 2         & -2            & 2         & -2            & 0         & 0         & 0         & 0         & -1        & 1         \\ \hline
    \end{tabular}
    \caption{\label{tb:tdls:z4} The fusion subcategory of the monster CFT inherited from the $\mb Z_4$ parafermion CFT. Here $\beta = (-1)^{1/4}$.}
\end{table}
Once again there is a Hecke operator relating the parafermion characters to those of the commutant, in this case $\hecke_{23}$. All lines besides powers of $\theta$ are non-invertible and so give novel defect McKay-Thompson series; $\theta$ implements the global symmetry corresponding to the $\mb Z_{4 A}$ conjugacy class of the monster.

\subsection{\boldmath $\qty(D_{3 \text A}, Fi_{23})$: $\mc W_{D_{3 \text A}}$ algebra}
Finally, we consider the monstralizer pair $\qty(D_{3 \text A}, Fi_{23})$. In the uplift to VOAs, $D_{3 A}$ becomes the chiral algebra $\mc W_{D_{3 \text A}}$, which has been studied in \cite{miyamoto_vertex_2003,sakuma_vertex_2003,lam_mckay_2005}. We take the monstralizing RCFT to be the theory with full chiral algebra $\mc W_{D_{3 \text A}} \otimes \overline{\mc W}_{D_{3 \text A}}$ and the diagonal modular invariant. Comparatively little is known about this VOA: in particular, we are not aware of any attempt to derive the full fusion category of TDLs in the theory. Fortunately, the characters of the theory may be decomposed into minimal model characters and, consequently, the modular $S$ and $T$ matrices may be derived \cite{bae_conformal_2021}. We may therefore obtain the Verlinde lines of the theory using Verlinde's formula \cite{verlinde_fusion_1988}, but there may be additional TDLs (preserving Virasoro but not the fully extended chiral algebra) not found by this method. The TDLs found in this way and their eigenvalues are given in table \ref*{tb:tdls:d3}.
\begin{table}
    \centering
    \begin{tabular}{| C | C C C C C C |}
    \hline
        h   & 0                     & \frac{1}{7}           & \frac{5}{7}           & \frac{2}{5}               & \frac{19}{35}             & \frac{4}{35}              \\ \hline
        1   & 1                     & 1                     & 1                     & 1                         & 1                         & 1                         \\
        A   & \alpha                & \beta                 & \gamma                & \alpha                    & \beta                     & \gamma                    \\
        B   & \gamma^{-1}           & \alpha^{-1}           & \beta^{-1}            & \gamma^{-1}               & \alpha^{-1}               & \beta^{-1}                \\
        W   & \varphi               & \varphi               & \varphi               & -\varphi^{-1}             & -\varphi^{-1}             & -\varphi^{-1}             \\
        A W & \alpha \varphi        & \beta \varphi         & \gamma \varphi        & -\alpha \varphi^{-1}      & -\beta \varphi^{-1}       & -\gamma \varphi^{-1}      \\
        B W & \gamma^{-1} \varphi   & \alpha^{-1} \varphi   & \beta^{-1} \varphi    & -\gamma^{-1} \varphi^{-1} & -\alpha^{-1} \varphi^{-1} & -\beta^{-1} \varphi^{-1}  \\ \hline
    \end{tabular}
    \caption{\label{tb:tdls:d3} The fusion subcategory of the monster CFT inherited from the $\mc W_{D_{3 \text A}}$ CFT. Here $\alpha = \frac{\cos(3 \pi/14)}{\sin(\pi/7)}$, $\beta = -2 \sin(\frac{3 \pi}{14})$, $\gamma = 2 \sin(\frac{\pi}{14})$, and $\varphi = \frac{1 + \sqrt{5}}{2}$, and the primaries are labeled by their (left-moving) conformal dimensions.}
\end{table}
All the Verlinde lines are non-invertible. This is the first case we have considered where the characters of the VOAs forming the monstralizer pair are not known to be related by a Hecke operator; however, the characters of the commutant theory solve a sixth-order modular linear differential equation \cite{bae_conformal_2021}, and one may pair them with the correct $\mc W_{D_{3 \text A}}$ characters by requiring that
\begin{equation}
    J(\tau) = \sum_i \chi^{(W_{D_{3 \text A}})}_i (\tau) \chi^{(Fi_{23})}_i (\tau).
\end{equation}
This allows us to compute $q$-expansions for the defect McKay-Thompson series.

\section{Invariance groups} \label{ss:invgrps}
Now that we have collected a number of defect McKay-Thompson series, we wish to find their modular invariance groups. We do so in two parts: first, we describe a general method to determine the invariance group in $\SL(2, \mb Z)$, and then consider whether the invariance group in $\SL(2, \mb R)$ might be larger.

\subsection{\boldmath In $\SL(2, \mb Z)$} \label{ss:sl2z}
Turning to the invariance groups in $\SL(2, \mb Z)$, we begin by expositing the method used by Lin and Shao to compute the invariance group of the defect McKay-Thompson series found in \cite{lin_duality_2021}. As we shall see, this method is not suited to the more general case, and so we develop an algorithm to determine the modular invariance group of any defect McKay-Thompson series in $\SL(2, \mb Z)$.

\subsubsection{Modular $S$ and $T$ matrices} \label{ss:modst}
In \cite{lin_duality_2021}, Lin and Shao showed that the defect McKay-Thompson series $Z^{\mc D}(\tau)$ (corresponding to the $\mc D$ line in the Ising category) is invariant under the action of the congruence subgroup of $\PSL(2, \mb Z)$ called $16 D^0$ in the notation of \cite{cummins_congruence_2003}. To see this, consider the larger congruence subgroup $\Gamma_0(8)$, where 
\begin{equation}
    \Gamma_0(N) = \qty{\mqty(a & b \\ c & d) \in \PSL(2, \mb Z) \ \bigg| \ c \equiv 0 \mod N}.
\end{equation}
Lin and Shao pointed out that $\Gamma_0(8)$ is generated by three matrices (say $T$, $G_1$, $G_2$) satisfying
\begin{equation}
    \begin{gathered} \label{eq:modst:zd}
    Z^{\mc D}(G_1 \cdot \tau) = -Z^{\mc D}(\tau), \qq{} Z^{\mc D}(G_2 \cdot \tau) = Z^{\mc D}(\tau) \\
    \qq{and} Z^{\mc D}(T \cdot \tau) = Z^{\mc D}(\tau + 1) = Z^{\mc D}(\tau).
    \end{gathered}
\end{equation}
Thus, $Z^{\mc D}(\tau)$ is invariant under any element of $\Gamma_0(8)$ which may be written as a word in $T$, $G_1$, and $G_2$ containing an even number of $G_1$. They went on to show that this subgroup is precisely $16 D^0$, a level 16, genus zero subgroup of $\PSL(2, \mb Z)$ \cite{cummins_congruence_2003}.

One might hope that Lin and Shao's procedure could be adapted to find the invariance groups in $SL(2, \mb Z)$ of other defect McKay-Thompson series; to that end, we note that 8 is the lowest value of $n$ for which the modular transformation $S T^n S$ fixes $Z^{\mc D}(\tau)$ up to a sign. This suggests a general procedure for finding the invariance groups: starting with a defect McKay-Thompson series $Z^{\mc L}(\tau)$, find the lowest value of $n$ such that
\begin{equation} \label{eq:modst:zl}
    Z^{\mc L}(S T^n S \cdot \tau) = \pm Z^{\mc L}(\tau).
\end{equation}
Then take a set of generators for the group $\Gamma_0(n)$ (say $\qty{G_i}$) and compute $Z^{\mc L}(G_i \cdot \tau)$ for each. Hopefully, one obtains relations like (\ref*{eq:modst:zd}), using which one can describe the invariance group in $\Gamma_0(n)$. Several difficulties prevent this. For one, this procedure does not guarantee that there are no other elements of $\SL(2, \mb Z)$ leaving $Z^{\mc D}(\tau)$ invariant (in other words, the invariance group of $Z^{\mc L}$ in $\SL(2, \mb Z)$ might be larger than its invariance group in $\Gamma_0(n)$); moreover, for more general defect McKay-Thompson series, the analogous relations to (\ref*{eq:modst:zd}) are complicated and do not lend themselves to a simple description of the invariance group in terms of restrictions on words in the generators. In the next section we describe a completely general method for computing the invariance group in $\SL(2, \mb Z)$ of a defect McKay-Thompson series.

\subsubsection{Coset permutations} \label{ss:cosets}
Our general method for finding the invariance groups in $\SL(2, \mb Z)$ begins with the following observation: from equations \ref*{eq:tdls:zl} and \ref*{eq:tdls:sb}, the defect McKay-Thompson series $Z^{\mc L}(\tau)$ is invariant under the modular transformation $M$ if
\begin{equation} \label{eq:cosets:dl}
    \rho(M)^\dag D_{\mc L} \rho(M) = D_{\mc L},
\end{equation}
where $\rho$ is the representation of $\SL(2, \mb Z)$ generated by the modular $S$ and $T$ matrices of $\mc A$. Moreover, this condition is \textit{necessary} provided that the characters $\chi_i^{(\mc A)}(\tau)$ appearing in the monstralizer decomposition (\ref*{eq:tdls:j}) are linearly independent. In all the cases under consideration in this work, the various characters either (i) have different leading powers of $q$ or (ii) are equal. In case (i), the characters must obviously be linearly independent. In case (ii) the Hecke images of the redundant characters will also be equal, which means we may represent the McKay-Thompson in an alternate way. Suppose that
\begin{equation}
    \chi_1^{(\mc A)}(\tau) = \chi_2^{(\mc A)}(\tau).
\end{equation}
then $\chi_1^{(\mc B)}(\tau)$ and $\chi_2^{(\mc B)}(\tau)$ are equal as well. Therefore, write 
\begin{equation}
    \vec{\tilde{\chi}}^{(\mc A)}(\tau) = \mqty(\sqrt{2} \chi_1^{(\mc A)}(\tau) \\ \chi_3^{(\mc A)}(\tau) \\ \chi_4^{(\mc A)}(\tau) \\ ...) \qq{and}
    \vec{\tilde{\chi}}^{(\mc B)}(\tau) = \mqty(\sqrt{2} \chi_1^{(\mc B)}(\tau) \\ \chi_3^{(\mc B)}(\tau) \\ \chi_4^{(\mc B)}(\tau) \\ ...).
\end{equation}
This furnishes our alternate representation of $Z^{\mc L}(\tau)$:
\begin{equation} \label{eq:cosets:zl}
    Z^{\mc L}(\tau) = \vec{\chi}^{(\mc A)}(\tau)^\transpose D_{\mc L} \vec{\chi}^{(\mc B)}(\tau) = \vec{\tilde{\chi}}^{(\mc A)}(\tau)^\transpose \tilde{D}_{\mc L} \vec{\tilde{\chi}}^{(\mc B)}(\tau)
\end{equation}
where
\begin{equation}
    \qty(\tilde{D}_{\mc L})_{i j} = \begin{cases}
        \frac{1}{2} \qty(\qty(D_{\mc L})_{1 1} + \qty(D_{\mc L})_{1 2} + \qty(D_{\mc L})_{2 1} + \qty(D_{\mc L})_{2 2}) & i = j = 1 \\
        \frac{1}{\sqrt{2}} \qty(\qty(D_{\mc L})_{1 (j + 1)} + \qty(D_{\mc L})_{2 (j + 1)}) & i = 1,\ j > 1 \\
        \frac{1}{\sqrt{2}} \qty(\qty(D_{\mc L})_{(i + 1) 1} + \qty(D_{\mc L})_{(i + 1) 2}) & i > 1,\ j = 1 \\
        \qty(D_{\mc L})_{(i + 1) (j + 1)} & i, j > 1
    \end{cases}.
\end{equation}
We pause to emphasize that equation \ref*{eq:cosets:zl} should be understood as a statement about modular functions, not about CFTs. The right-hand side of (\ref*{eq:cosets:zl}) does not correspond to a true decomposition into subVOA representations: we have not constructed a new CFT with characters $\vec{\tilde{\chi}}^{(\mc A)}$. Rather, (\ref*{eq:cosets:zl}) is merely a convenient way of rewriting $\mc Z^{\mc L}(\tau)$ in terms of linearly independent objects. Nonetheless, one may check that $\vec{\tilde{\chi}}^{(\mc A)}(\tau)$ and $\vec{\tilde{\chi}}^{(\mc B)}(\tau)$ in fact transform linearly under modular transformations, with ``modified'' $S$ and $T$-matrices
\begin{equation}
\begin{gathered}
    \tilde{T}^{(\mc A)} = \diag(T^{(\mc A)}_{1 1}, T^{(\mc A)}_{3 3}, T^{(\mc A)}_{4 4}, ...) \\
    \tilde{S}^{(\mc A)}_{i j} = \begin{cases}
        \frac{1}{4} \qty(S^{(\mc A)}_{1 1} + S^{(\mc A)}_{1 2} + S^{(\mc A)}_{2 1} + S^{(\mc A)}_{2 2}) & i = j = 1 \\
        \frac{1}{2} \qty(S^{(\mc A)}_{1 (j + 1)} + S^{(\mc A)}_{2 (j + 1)}) & i = 1,\ j > 1 \\
        \frac{1}{2} \qty(S^{(\mc A)}_{(i + 1) 1} + S^{(\mc A)}_{(i + 1) 2}) & i > 1,\ j = 1 \\
        S^{(\mc A)}_{(i + 1) (j + 1)} & i, j > 1
    \end{cases} \\
    \tilde{T}^{(\mc B)} = T^{(\mc A) *} \qq{and} \tilde{S}^{(\mc B)} = S^{(\mc A) *}.
\end{gathered}
\end{equation}
Thus, the problem of finding the modular invariance group of $D_{\mc L}$ transforming in the representation generated by $S^{(\mc A)}$ and $T^{(\mc A)}$ is identical to that of finding the modular invariance group of $\tilde{D}_{\mc L}$ transforming in the representation generated by $\tilde{S}^{(\mc A)}$ and $\tilde{T}^{(\mc A)}$. This procedure allows us to remove the redundant characters, leaving only a linearly independent list behind. Going forward we will have done so, and it may be assumed that the different elements of $\vec{\chi}^{(\mc A)}(\tau)$ and $\vec{\chi}^{(\mc B)}(\tau)$ are linearly independent, in particular having different leading powers of $q$ (though again we note that these elements are no longer necessarily characters of any RCFT, and should be thought of simply as modular functions which we have chosen for convenience).

Next, let $\Gamma$ be the invariance group of $Z^{\mc L}(\tau)$ in $\SL(2, \mb Z)$. We make the following claim: $\SL(2, \mb Z)$ conjugates of $D_{\mc L}$ (i.e. matrices of the form $\rho(M)^\dag D_{\mc L} \rho(M)$, $M \in \SL(2, \mb Z)$) are in one-to-one correspondence with the right cosets of $\Gamma$. To see this, let $G_1$ and $G_2$ be two elements of $SL(2, \mb Z)$. Note that
\begin{equation}
    \begin{aligned}
        {}& \rho(G_1)^\dag D_{\mc L} \rho(G_1) = \rho(G_2)^\dag D_{\mc L} \rho(G_2) \\
        \iff {}& D_{\mc L} = \rho\qty(G_2 G_1^{-1})^\dag D_{\mc L} \rho\qty(G_2 G_1^{-1}) \\
        \iff {}& G_2 G_1^{-1} \in \Gamma \qq{(by equation \ref*{eq:cosets:dl})} \\
        \iff {}& G_2 \in \Gamma G_1.
    \end{aligned}
\end{equation}
Thus, the $G_{1, 2}$-conjugates of $D$ are equal if and only if $G_1$ and $G_2$ lie in the same right coset of $\Gamma$, so distinct matrices of the form $\rho(M)^\dag D_{\mc L} \rho(M)$ are indeed in one-to-one correspondence with right cosets of $\Gamma$. Moreover, the permutation of the right cosets induced by a given element of $\Gamma$ acting from the right is the same as the permutation of the $\Gamma$-conjugates of $D$ induced by conjugation by that element.

The utility of this correspondence is in the following fact: a finite-index subgroup of a finitely generated group is entirely determined by the action of the generators on its right cosets, meaning the permutation of the right cosets induced by the action of each generator from the right. This means that, if we know the possible $\SL(2, \mb Z)$ conjugates of $D_{\mc L}$, we can find the invariance group by conjugating each one by $\rho(S)$ and $\rho(T)$, and recording how each generator permutes the conjugates. To see this, begin with $D_{\mc L}$ and successively conjugate with $\rho(S)$ and $\rho(T)$, building a directed graph with two edges exiting each node, until the graph closes on itself. This is the Cayley graph corresponding to the action of $SL(2, \mb Z)$ on the cosets; the graph will contain as many nodes as there are distinct $\SL(2, \mb Z)$ conjugates of $D_{\mc L}$ (which is the same as the number of cosets of $\Gamma$, a.k.a. the index of $\Gamma$ in $\SL(2, \mb Z)$). Then, find a generating set for the fundamental group of the directed graph. Each of the elements of the generating set is a sequence of edges labeled with generators of $\SL(2, \mb Z)$ and so may be translated into an element of $\SL(2, \mb Z)$; these elements then generate $\Gamma$. In practice, once we have built the Cayley graph, we may number the nodes arbitrarily and find the permutation of cosets induced by $S$ (resp. $T$) by following the chain of consecutive $S$ (resp. $T$) edges. Given the permutations, the Sage mathematics system provides a built-in method for determining the invariance group \cite{sagemath}. An example graph is shown in figure \ref*{fig:cosets:graph}.
\begin{figure}[]
    \centering
    \begin{tikzpicture}[
        node distance = 7mm and 9mm,
        graphNode/.style={circle, draw, thick, minimum size=7mm, inner sep=0mm, align=center, text width=7mm},
        graphEdge/.style={->, > = latex', thick},
        graphLabel/.style={midway, above, sloped, draw=none}
        ]
        \node[graphNode] (1) {1};
        \node[graphNode] (2) [below=of 1] {2};
        \node[graphNode] (3) [right=of 2] {3};
        \node[graphNode] (5) [right=of 3] {5};
        \node[graphNode] (6) [below=of 5] {6};
        \node[graphNode] (7) [below=of 6] {7};
        \node[graphNode] (8) [below=of 7] {8};
        \node[graphNode] (4) [left=of 8] {4};
        \node[graphNode] (9) [right=of 5] {9};
        \node[graphNode] (10) [right=of 8] {10};
        \node[graphNode] (11) [right=of 9] {11};
        \node[graphNode] (12) [below=of 11] {12};
        \node[graphNode] (13) [below=of 12] {13};
        \node[graphNode] (14) [below=of 13] {14};
        \node[graphNode] (15) [right=of 11] {15};
        \node[graphNode] (16) [right=of 14] {16};
        \node[graphNode] (17) [right=of 15] {17};
        \node[graphNode] (18) [below=of 17] {18};
        \node[graphNode] (19) [below=of 18] {19};
        \node[graphNode] (20) [right=of 16] {20};
        \node[graphNode] (21) [right=of 17] {21};
        \node[graphNode] (22) [right=of 20] {22};
        \node[graphNode] (23) [right=of 21] {23};
        \node[graphNode] (24) [above=of 23] {24};
        \node[draw=none] (25) [above=of 5] {};
        \node[draw=none] (26) [above=of 9] {};
        \node[draw=none] (27) [above=of 15] {};
        \node[draw=none] (28) [above=of 17] {};
        \draw[graphEdge] (1) to [bend right] (2);
        \draw[graphEdge] (1) to [loop right] (1) [dashed];
        \draw[graphEdge] (2) to [bend right] (1);
        \draw[graphEdge] (2) to [bend left] (3) [dashed];
        \draw[graphEdge] (3) to [bend right=15] (4);
        \draw[graphEdge] (3) to [bend left] (5) [dashed];
        \draw[graphEdge] (4) to [bend right=15] (3);
        \draw[graphEdge] (4) to [bend left] (2) [dashed];
        \draw[graphEdge] (5) to [bend right] (6);
        \draw[graphEdge] (5) to [bend left] (9) [dashed];
        \draw[graphEdge] (6) to [bend right] (5);
        \draw[graphEdge] (6) to [bend right] (7) [dashed];
        \draw[graphEdge] (7) to [bend right] (8);
        \draw[graphEdge] (7) to [bend right] (6) [dashed];
        \draw[graphEdge] (8) to [bend right] (7);
        \draw[graphEdge] (8) to [bend left] (4) [dashed];
        \draw[graphEdge] (9) to [bend right=15] (10);
        \draw[graphEdge] (9) to [bend left] (11) [dashed];
        \draw[graphEdge] (10) to [bend right=15] (9);
        \draw[graphEdge] (10) to [bend left] (8) [dashed];
        \draw[graphEdge] (11) to [bend right] (12);
        \draw[graphEdge] (11) to [bend left] (15) [dashed];
        \draw[graphEdge] (12) to [bend right] (11);
        \draw[graphEdge] (12) to [bend right] (13) [dashed];
        \draw[graphEdge] (13) to [bend right] (14);
        \draw[graphEdge] (13) to [bend right] (12) [dashed];
        \draw[graphEdge] (14) to [bend right] (13);
        \draw[graphEdge] (14) to [bend left] (10) [dashed];
        \draw[graphEdge] (15) to [bend right=15] (16);
        \draw[graphEdge] (15) to [bend left] (17) [dashed];
        \draw[graphEdge] (16) to [bend right=15] (15);
        \draw[graphEdge] (16) to [bend left] (14) [dashed];
        \draw[graphEdge] (17) to [bend right] (18);
        \draw[graphEdge] (17) to [bend left] (21) [dashed];
        \draw[graphEdge] (18) to [bend right] (17);
        \draw[graphEdge] (18) to [bend right] (19) [dashed];
        \draw[graphEdge] (19) to [bend right] (20);
        \draw[graphEdge] (19) to [bend right] (18) [dashed];
        \draw[graphEdge] (20) to [bend right] (19);
        \draw[graphEdge] (20) to [bend left] (16) [dashed];
        \draw[graphEdge] (21) to [bend right=15] (22);
        \draw[graphEdge] (21) to [bend left] (23) [dashed];
        \draw[graphEdge] (22) to [bend right=15] (21);
        \draw[graphEdge] (22) to [bend left] (20) [dashed];
        \draw[graphEdge] (23) to [bend right] (24);
        \draw[graphEdge] (23) to [bend left] (22) [dashed];
        \draw[graphEdge] (24) to [bend right] (23);
        \draw[graphEdge] (24) to [loop right] (24) [dashed];
        \draw[graphEdge] (25) to node[graphLabel] {$S$} (26);
        \draw[graphEdge] (27) to node[graphLabel] {$T$} (28) [dashed];
        \end{tikzpicture}
    \caption{An example of the directed graphs described in section \ref*{ss:cosets}. This graph corresponds to the $\mc D$ line in the Ising subcategory, which was studied by Lin and Shao in \cite{lin_duality_2021} and shown to be invariant under the congruence subgroup $16 D^0$ in the notation of \cite{cummins_congruence_2003}. Each node respresents a distinct $\SL(2, \mb Z)$ conjugate of the diagonal matrix $D_{\mc D}$ (the node labeled 1 represents $D_{\mc D}$ itself). Solid lines represent conjugation by the $\SL(2, \mb Z)$ generator $S$, while dashed lines represent $T$. In this case, the permutations induced by the various generators would be (in cycle notation): $S \to \qty(1\ 2)(3\ 4)(5\ 6)(7\ 8)(9\ 10)(11\ 12)(13\ 14)(15\ 16)(17\ 18)(19\ 20)(21\ 22)(23\ 24)$, $T \to (2\ 3\ 5\ 9\ 11\ 15\ 17\ 21\ 23\ 22\ 20\ 16\ 14\ 10\ 8\ 4)(6\ 7)(12\ 13)(18\ 19)$.}
    \label{fig:cosets:graph}
\end{figure}

Finally, we note that the permutations induced by each generator uniquely fix a subgroup of $SL(2, \mb Z)$ irrespective of any relabeling of the right cosets. For a given defect $\mc L$, the twined, twisted, and twisted + twined partition functions are simply related by acting by $S$ and $T$, or equivalently by conjugating $D_{\mc L}$ by $\rho(S)$ and $\rho(T)$, which amounts to relabeling the nodes of the directed graph. Therefore the twined, twisted, and twisted + twined partition functions share the same invariance group in $SL(2, \mb Z)$.

\subsection{\boldmath In $\SL(2, \mb R)$: Atkin-Lehner involutions \label{ss:sl2r}}
Some McKay-Thompson series have invariance groups in $SL(2, \mb Z)$ which are not genus zero. For these series, one must include elements of $SL(2, \mb R)$ in order to obtain a genus zero invariance group \cite{borcherds_monstrous_1992,paquette_monstrous_2016}. One might wonder if defect McKay-Thompson series are the same way: perhaps one must consider $\SL(2, \mb R)$ transformations to obtain a genus zero invariance group. On the other hand, if the invariance group in $\SL(2, \mb Z)$ is already genus zero, the invariance group in $\SL(2, \mb R)$ must be as well, since the genus of a subgroup is always greater than or equal to the genus of the supergroup \cite{cummins_congruence_2003}. This suggests that we check, for each of our TDLs $\mc L$ whose invariace group in $SL(2, \mb Z)$ is not genus zero, whether there are (non-integral) elements of $\SL(2, \mb R)$ which leave $Z^{\mc L}(\tau)$ invariant.

In general, since $\SL(2, \mb R)$ is not finitely generated, this is an intractable problem; that said, there is a particular class of $SL(2, \mb R)$ transformations which we are well-motivated to consider: the Atkin-Lehner involutions. Conway and Norton \cite{conway_monstrous_1979} argued that the invariance groups of the McKay-Thompson series have the form $\Gamma_0(N) + e, f, g,...$, where $N$ is the level of the invariance group as a congruence subgroup of $\SL(2, \mb Z)$, and $+ e, f, g...$ signifies that we may have to adjoin certain cosets of $\Gamma_0(N)$, labeled by integers $e, f, g$ which exactly divide $N$ ($e \mathrel\Vert N$ means $e \mid N$ and $(e, N/e)$ are relatively prime). Since we may determine the level of $\Gamma$ even if we only know the invariance group in $SL(2, \mb Z)$, we may assume for the moment that the defect McKay-Thompson series behave similarly, and explicitly check whether any of these cosets fix $Z_{\mc L}(\tau)$. If not, we know that either the invariance group in $SL(2, \mb R)$ does not extend that in $\SL(2, \mb Z)$, or the defect McKay-Thompson series have invariance groups in $SL(2, \mb R)$ of a different form from those of the ``normal'' McKay-Thompson series.

To perform the checks, we need formulae for the characters appearing in the monstralizer decompositions (\ref*{eq:tdls:j}). Fortunately, in nearly all of the cases discussed in section \ref*{ss:tdls}, one set of characters (those of the ``small'' RCFT) is given in terms of minimal model characters, and the other set is given by the action of a vector-valued Hecke operator on the first. Only in the case with monstralizer pair $\qty(D_{3 \text A}, Fi_{23})$ are the characters related by a modular linear differential equation without a (known) solution in terms of Hecke operators. In all other cases, we may therefore express $Z_{\mc L}(\tau)$ as an infinite sum. Next, the $\SL(2, \mb R)$-transformed characters are computed from the untransformed ones by Poisson resummation, allowing us to check whether a particular transformation fixes $Z_{\mc L}(\tau)$. For each series which is not genus zero, we may obtain the level using our method for finding the invariance group in $\SL(2, \mb Z)$, and then check each of the associated Atkin-Lehner involutions to see if the series is invariant. Doing so, we do not find that any of the Atkin-Lehner involutions extend the invariance groups in any case (apart from $D_{3 A}$, which we cannot check with our methods).

\section{Results} \label{ss:results}
In this section we present our main results: the invariance groups of all of the (defect) McKay-Thompson series described in section \ref*{ss:tdls}. Table \ref*{tb:results:invgrps} gives, for each of the TDLs studied,
\begin{enumerate}
    \item the TDL's label (cf. section \ref*{ss:tdls}),
    \item whether the TDL is invertible (and so whether its McKay-Thompson series is of the traditional or ``defect'' type),
    \item the level of the invariance group as a congruence subgroup of $\PSL(2, \mb Z)$,
    \item the genus of the invariance group,
    \item the index of the invariance group in $SL(2, \mb Z)$, and
    \item the label of the invariance group, in the notation of \cite{cummins_congruence_2003}, if available. ($*$) means that the group has genus greater than 3, in which case \cite{cummins_congruence_2003} does not give a label for it. For these groups, generators are given in appendix. 
\end{enumerate}
\begin{table}
    \centering
    \begin{tabular}{| C | c | c | c | c | C |}
        \hline
        \text{TDL}  & Invertible?   & Level & Genus & Index & \text{Label}  \\ \hline
        \multicolumn{6}{| c |}{Ising}                                       \\ \hline
        \eta        & Yes           & 2     & 0     & 3     & 2 B^0         \\
        \mc D       & No            & 16    & 0     & 24    & 16 D^0        \\ \hline
        \multicolumn{6}{| c |}{Three-state Potts}                           \\ \hline
        \eta        & Yes           & 3     & 0     & 4     & 3 B^0         \\
        \eta^2      & Yes           & 3     & 0     & 4     & 3 B^0         \\
        W           & No            & 5     & 0     & 12    & 5 D^0         \\
        \eta W      & No            & 15    & 1     & 48    & 15 G^1        \\
        \eta^2 W    & No            & 15    & 1     & 48    & 15 G^1        \\
        N           & No            & 24    & 0     & 48    & 24 B^0        \\
        W N         & No            & 120   & 29    & 576   & *        \\ \hline
        \multicolumn{6}{| c |}{$\mb Z_4$ parafermion}                       \\ \hline
        \theta      & Yes           & 4     & 0     & 6     & 4 B^0         \\
        \theta^2    & Yes           & 2     & 0     & 3     & 2 B^0         \\
        \theta^3    & Yes           & 4     & 0     & 6     & 4 B^0         \\
        X           & No            & 48    & 7     & 192   & *             \\
        \theta X    & No            & 48    & 7     & 192   & *             \\
        \theta^2 X  & No            & 48    & 7     & 192   & *             \\
        \theta^3 X  & No            & 48    & 7     & 192   & *             \\
        Y           & No            & 6     & 0     & 12    & 6 F^0         \\
        \theta Y    & No            & 12    & 0     & 24    & 12 E^0        \\ \hline
        \multicolumn{6}{| c |}{$\mc W_{D_{3 \text A}}$}                     \\ \hline
        A           & No            & 7     & 0     & 24    & 7 E^0         \\
        B           & No            & 7     & 0     & 24    & 7 E^0         \\
        W           & No            & 5     & 0     & 12    & 5 D^0         \\
        A W         & No            & 35    & 13    & 288   & *             \\
        B W         & No            & 35    & 13    & 288   & *             \\ \hline
    \end{tabular}
    \caption{Our main results: the invariance groups associated to the TDLs in the Ising, three-state Potts, $\mb Z_4$ parafermion, and $\mc W_{D_{3 \text A}}$ subcategories of the monster CFT.}
    \label{tb:results:invgrps}
\end{table}
The invariance groups for the invertible lines are genus zero as expected, but many of the invariance groups for the defect McKay-Thompson series are seen to have nonzero genus.

For reference, we also provide $q$ expansions for selected defect McKay-Thompson series.
\footnote{These $q$ expansions were constructed using the relationships between the characters of the communtant pairs described in section \ref{ss:tdls}. Many of these relationships are cataloged in \cite{bae_conformal_2021} -- however, some of the $q$-expansions for the commutant characters in that paper contain typos; to recreate these results the vector-valued Hecke operators and MLDEs should be used to recover the commutant characters as appropriate.}
For the Ising subcategory, the defect McKay-Thompson series associated to the Kramers-Wannier defect $\mc D$ is
\begin{equation}
    Z^{\mc D}(q) = \sqrt{2} \qty(\frac{1}{q} + 91886 q + 8498776 q^2 + 301112552 q^3 + ...)
\end{equation}
which agrees with the $q$-expansion presented in \cite{lin_duality_2021}. All other lines in the subcategory are invertible and so correspond to ``typical'' McKay-Thompson series.

For the three-state Potts subcategory, the defect McKay-Thompson series associated to the line $W$ is
\begin{equation}
    \begin{aligned}
        Z^W(q) ={}& \frac{1}{2} \qty(\frac{1}{q} + 196884 q + 21493760 q^2 + 864299970 q^3 + ...) \\
        {}& \qq{} + \frac{\sqrt{5}}{2} \qty(\frac{1}{q} - 78794 q - 9307840 q^2 - 381508370 q^3 + ...).
    \end{aligned}
\end{equation}

For the $\mb Z_4$ parafermion subcategory, the defect McKay-Thompson series associated to the line $Y$ is
\begin{equation}
    Z^Y(q) = \frac{2}{q} + 181547 q + 19336938 q^2 + ... .
\end{equation}

For the $\mc W_{D_{3 A}}$ subcategory, the defect McKay-Thompson series associated to the line $A$ is
\begin{equation}
    \begin{aligned}
        Z^A(q) ={}& \cos(\frac{3 \pi}{14}) \csc(\frac{\pi}{7}) \qty(\frac{1}{q} - 9618984 q^2 + ...) \\
        {}& \qq{} + (36570 q + 5104206 q^2 + 229935232 q^3 + ...) \\
        {}& \qq{} + \frac{1}{2} \csc(\frac{\pi}{14}) \qty(176130 q + 7900063 q^2 + 663713370 q^3 + ...) \\
        {}& \qq{} + \cos(\frac{\pi}{7}) \csc(\frac{\pi}{14}) \qty(-88362 q - 384310822 q^3 + ...).
    \end{aligned}
\end{equation}

\section{Conclusions} \label{ss:conclusions}
Using the monstralizer pairs of CFTs cataloged in \cite{bae_conformal_2021}, we have found a number of topological defect lines (TDLs) appearing in the monster CFT. For each of these we have computed the modular invariance group of the associated defect McKay-Thompson series in $\SL(2, \mb Z)$, and given (partial) evidence that the invariance group in $SL(2, \mb R)$ is not larger. Thus, we conclude that the invariance groups of defect McKay-Thompson series either
\begin{itemize}
    \item do not generically satisfy the genus-zero property, or
    \item differ fundamentally from the invariance groups of ``normal'' McKay-Thompson series in that they contain non-integral elements of $SL(2, \mb R)$ other than Atkin-Lehner involutions.
\end{itemize}
In the future, we may wish to understand whether the non-invertible lines which are genus zero fit some pattern, as well as how these results fit into the physical interpretation of the genus zero property proposed in \cite{paquette_monstrous_2016, paquette_bps_2017}.

\section{Acknowledgements}
This work was supported by the National Science Foundation Grant PHY-2310635. JH thanks S.-H.~Shao and G.~Moore for helpful conversations. We thank one of the referees for several helpful remarks. 
\appendix
\section{Generators for the invariance groups of high genus}
Here we present generators for the invariance groups in table \ref*{tb:results:invgrps} of genus greater than 3, since the information provided in the table is not sufficient to uniquely determine them.

The invariance group of the TDL of the $\mb Z_3$ parafermion subcategory labeled $W N$ is generated by
\begin{equation}
    \begin{aligned}
    \big\{{}& \qty(\smqty{1 & 1 \\ 0 & 1}), \qty(\smqty{109 & -5 \\ 240 & -11}), \qty(\smqty{61 & -3 \\ 1200 & -59}), \qty(\smqty{641 & -33 \\ 2700 & -139}), \qty(\smqty{1019 & -53 \\ 3480 & -181}), \qty(\smqty{391 & -21 \\ 540 & -29}), \qty(\smqty{209 & -12 \\ 540 & -31}), \qty(\smqty{989 & -60 \\ 1500 & -91}), \\
    {}& \qty(\smqty{221 & -14 \\ 300 & -19}), \qty(\smqty{191 & -13 \\ 720 & -49}), \qty(\smqty{541 & -38 \\ 840 & -59}), \qty(\smqty{149 & -11 \\ 420 & -31}), \qty(\smqty{1261 & -99 \\ 2280 & -179}), \qty(\smqty{1499 & -119 \\ 2280 & -181}), \qty(\smqty{61 & -5 \\ 720 & -59}), \\
    {}& \qty(\smqty{439 & -38 \\ 1860 & -161}), \qty(\smqty{581 & -51 \\ 900 & -79}), \qty(\smqty{109 & -10 \\ 120 & -11}), \qty(\smqty{161 & -17 \\ 180 & -19}), \qty(\smqty{631 & -72 \\ 780 & -89}), \qty(\smqty{1379 & -160 \\ 1560 & -181}), \qty(\smqty{421 & -49 \\ 3600 & -419}), \\
    {}& \qty(\smqty{401 & -48 \\ 660 & -79}), \qty(\smqty{299 & -38 \\ 480 & -61}), \qty(\smqty{139 & -18 \\ 780 & -101}), \qty(\smqty{581 & -76 \\ 1980 & -259}), \qty(\smqty{659 & -87 \\ 2280 & -301}), \qty(\smqty{139 & -19 \\ 300 & -41}), \qty(\smqty{169 & -25 \\ 480 & -71}), \\
    {}& \qty(\smqty{251 & -38 \\ 720 & -109}), \qty(\smqty{31 & -5 \\ 180 & -29}), \qty(\smqty{139 & -26 \\ 540 & -101}), \qty(\smqty{101 & -19 \\ 420 & -79}), \qty(\smqty{691 & -132 \\ 780 & -149}), \qty(\smqty{539 & -104 \\ 1560 & -301}), \qty(\smqty{619 & -120 \\ 1140 & -221}), \\
    {}& \qty(\smqty{289 & -57 \\ 360 & -71}), \qty(\smqty{191 & -39 \\ 240 & -49}), \qty(\smqty{349 & -73 \\ 1200 & -251}), \qty(\smqty{151 & -32 \\ 420 & -89}), \qty(\smqty{659 & -142 \\ 840 & -181}), \qty(\smqty{781 & -169 \\ 3600 & -779}), \qty(\smqty{229 & -50 \\ 600 & -131}), \\
    {}& \qty(\smqty{79 & -18 \\ 180 & -41}), \qty(\smqty{211 & -49 \\ 900 & -209}), \qty(\smqty{1171 & -277 \\ 3420 & -809}), \qty(\smqty{409 & -106 \\ 1200 & -311}), \qty(\smqty{991 & -257 \\ 1500 & -389}), \qty(\smqty{589 & -154 \\ 960 & -251}), \qty(\smqty{571 & -153 \\ 780 & -209}), \\
    {}& \qty(\smqty{329 & -89 \\ 780 & -211}), \qty(\smqty{559 & -154 \\ 1020 & -281}), \qty(\smqty{661 & -183 \\ 1080 & -299}), \qty(\smqty{329 & -92 \\ 540 & -151}), \qty(\smqty{991 & -286 \\ 1140 & -329}), \qty(\smqty{1259 & -365 \\ 2280 & -661}), \qty(\smqty{1139 & -333 \\ 1440 & -421}), \\
    {}& \qty(\smqty{169 & -50 \\ 240 & -71}), \qty(\smqty{41 & -13 \\ 60 & -19}), \qty(\smqty{4921 & -1681 \\ 14400 & -4919}), \qty(\smqty{3001 & -1026 \\ 4560 & -1559}), \qty(\smqty{449 & -154 \\ 1140 & -391}), \qty(\smqty{311 & -107 \\ 840 & -289}), \qty(\smqty{449 & -156 \\ 780 & -271}), \\
    {}& \qty(\smqty{331 & -121 \\ 900 & -329}), \qty(\smqty{419 & -158 \\ 480 & -181}), \qty(\smqty{1009 & -383 \\ 2400 & -911}), \qty(\smqty{1301 & -495 \\ 2100 & -799}), \qty(\smqty{389 & -152 \\ 540 & -211}), \qty(\smqty{581 & -228 \\ 660 & -259}), \qty(\smqty{71 & -29 \\ 120 & -49}), \\
    {}& \qty(\smqty{349 & -146 \\ 600 & -251}), \qty(\smqty{1861 & -781 \\ 3000 & -1259}), \qty(\smqty{509 & -216 \\ 780 & -331}), \qty(\smqty{919 & -420 \\ 1140 & -521}), \qty(\smqty{649 & -298 \\ 1200 & -551}), \qty(\smqty{2971 & -1366 \\ 4500 & -2069}), \qty(\smqty{1979 & -911 \\ 3000 & -1381}), \\
    {}& \qty(\smqty{31 & -15 \\ 60 & -29}), \qty(\smqty{259 & -139 \\ 300 & -161}), \qty(\smqty{661 & -363 \\ 1200 & -659}), \qty(\smqty{869 & -481 \\ 1140 & -631}), \qty(\smqty{139 & -78 \\ 180 & -101}), \qty(\smqty{569 & -329 \\ 780 & -451}), \qty(\smqty{749 & -454 \\ 1140 & -691}), \\
    {}& \qty(\smqty{469 & -290 \\ 600 & -371}), \qty(\smqty{551 & -347 \\ 840 & -529}), \qty(\smqty{571 & -361 \\ 900 & -569}), \qty(\smqty{331 & -212 \\ 420 & -269}), \qty(\smqty{409 & -265 \\ 480 & -311}), \qty(\smqty{611 & -398 \\ 720 & -469}), \qty(\smqty{1259 & -824 \\ 1560 & -1021}), \\
    {}& \qty(\smqty{929 & -612 \\ 1140 & -751}), \qty(\smqty{341 & -259 \\ 420 & -319}), \qty(\smqty{691 & -529 \\ 900 & -689}), \qty(\smqty{679 & -558 \\ 780 & -641}), \qty(\smqty{151 & -125 \\ 180 & -149}), \qty(\smqty{-1 & 0 \\ 0 & -1})
    \big\}.
    \end{aligned}
\end{equation}
The invariance groups of the TDLs of the $\mb Z_4$ parafermion subcategory labeled $X$, $\theta X$, $\theta^2 X$, and $\theta^3 X$ are identical, and are generated by \\
\begin{equation}
    \begin{aligned}
    \big\{& \qty(\smqty{1 & 1 \\ 0 & 1}),\ \qty(\smqty{43 & -3 \\ 72 & -5}),\ \qty(\smqty{25 & -2 \\ 288 & -23}),\ \qty(\smqty{73 & -7 \\ 240 & -23}),\ \qty(\smqty{605 & -62 \\ 888 & -91}), \qty(\smqty{241 & -25 \\ 2304 & -239}), \qty(\smqty{265 & -28 \\ 672 & -71}), \\
    {}& \qty(\smqty{53 & -6 \\ 168 & -19}),\ \qty(\smqty{67 & -9 \\ 216 & -29}),\ \qty(\smqty{163 & -24 \\ 360 & -53}), \qty(\smqty{667 & -101 \\ 984 & -149}),\ \qty(\smqty{25 & -4 \\ 144 & -23}), \qty(\smqty{265 & -49 \\ 384 & -71}),\ \qty(\smqty{263 & -50 \\ 384 & -73}), \\
    {}& \qty(\smqty{19 & -4 \\ 24 & -5}),\ \qty(\smqty{53 & -14 \\ 72 & -19}),\ \qty(\smqty{265 & -73 \\ 432 & -119}),\ \qty(\smqty{139 & -39 \\ 360 & -101}),\ \qty(\smqty{169 & -49 \\ 576 & -167}),\ \qty(\smqty{169 & -50 \\ 240 & -71}),\ \qty(\smqty{53 & -19 \\ 120 & -43}), \\
    {}& \qty(\smqty{73 & -27 \\ 192 & -71}),\ \qty(\smqty{307 & -120 \\ 504 & -197}),\ \qty(\smqty{913 & -358 \\ 1344 & -527}),\ \qty(\smqty{409 & -161 \\ 912 & -359}),\ \qty(\smqty{361 & -147 \\ 528 & -215}),\ \qty(\smqty{317 & -130 \\ 456 & -187}), \\
    {}& \qty(\smqty{121 & -50 \\ 288 & -119}),\ \qty(\smqty{265 & -121 \\ 576 & -263}),\ \qty(\smqty{25 & -12 \\ 48 & -23}),\ \qty(\smqty{211 & -117 \\ 312 & -173}),\ \qty(\smqty{77 & -43 \\ 120 & -67}),\ \qty(\smqty{121 & -75 \\ 192 & -119}),\ \qty(\smqty{-1 & 0 \\ 0 & -1}) \big\}.
    \end{aligned}
\end{equation}
The invariance groups associated to the TDLs $A W$ and $B W$ in the $\mc W_{D_{3 \text A}}$ subcategory are identical, and are generated by 
\begin{equation}
    \begin{aligned}
    \big\{{}& \qty(\smqty{1 & 1 \\ 0 & 1}),\ \qty(\smqty{29 & -1 \\ 175 & -6}),\ \qty(\smqty{111 & -4 \\ 805 & -29}),\ \qty(\smqty{531 & -20 \\ 770 & -29}),\ \qty(\smqty{76 & -3 \\ 735 & -29}),\ \qty(\smqty{559 & -23 \\ 875 & -36}),\ \qty(\smqty{71 & -3 \\ 805 & -34}),\ \qty(\smqty{321 & -14 \\ 665 & -29}), \\
    {}& \qty(\smqty{281 & -13 \\ 735 & -34}),\ \qty(\smqty{41 & -2 \\ 595 & -29}),\ \qty(\smqty{176 & -9 \\ 665 & -34}),\ \qty(\smqty{314 & -17 \\ 665 & -36}),\ \qty(\smqty{71 & -4 \\ 1225 & -69}),\ \qty(\smqty{314 & -19 \\ 595 & -36}),\ \qty(\smqty{251 & -16 \\ 455 & -29}), \\
    {}& \qty(\smqty{146 & -11 \\ 385 & -29}),\ \qty(\smqty{139 & -11 \\ 455 & -36}),\ \qty(\smqty{106 & -9 \\ 1225 & -104}),\ \qty(\smqty{139 & -13 \\ 385 & -36}),\ \qty(\smqty{426 & -41 \\ 665 & -64}),\ \qty(\smqty{169 & -18 \\ 385 & -41}),\ \qty(\smqty{76 & -9 \\ 245 & -29}), \\
    {}& \qty(\smqty{169 & -22 \\ 315 & -41}),\ \qty(\smqty{244 & -33 \\ 525 & -71}),\ \qty(\smqty{29 & -5 \\ 35 & -6}),\ \qty(\smqty{111 & -23 \\ 140 & -29}),\ \qty(\smqty{216 & -47 \\ 455 & -99}),\ \qty(\smqty{244 & -55 \\ 315 & -71}),\ \qty(\smqty{281 & -64 \\ 1225 & -279}), \\
    {}& \qty(\smqty{64 & -15 \\ 175 & -41}),\ \qty(\smqty{111 & -29 \\ 245 & -64}),\ \qty(\smqty{211 & -57 \\ 385 & -104}),\ \qty(\smqty{76 & -21 \\ 105 & -29}),\ \qty(\smqty{99 & -29 \\ 140 & -41}),\ \qty(\smqty{386 & -117 \\ 805 & -244}),\ \qty(\smqty{386 & -121 \\ 1225 & -384}), \\
    {}& \qty(\smqty{71 & -23 \\ 105 & -34}),\ \qty(\smqty{181 & -63 \\ 385 & -134}),\ \qty(\smqty{316 & -111 \\ 595 & -209}),\ \qty(\smqty{181 & -64 \\ 280 & -99}),\ \qty(\smqty{64 & -25 \\ 105 & -41}),\ \qty(\smqty{41 & -17 \\ 70 & -29}),\ \qty(\smqty{99 & -43 \\ 175 & -76}), \\
    {}& \qty(\smqty{561 & -256 \\ 1225 & -559}),\ \qty(\smqty{246 & -113 \\ 455 & -209}),\ \qty(\smqty{274 & -131 \\ 525 & -251}),\ \qty(\smqty{414 & -199 \\ 595 & -286}),\ \qty(\smqty{356 & -187 \\ 455 & -239}),\ \qty(\smqty{134 & -85 \\ 175 & -111}),\ \qty(\smqty{-1 & 0 \\ 0 & -1}) \big\}.
    \end{aligned}
\end{equation}

\clearpage

\bibliographystyle{JHEP}
\bibliography{bib.bib}

\providecommand{\href}[2]{#2}\begingroup\raggedright\begin{thebibliography}{10}

\bibitem{conway_monstrous_1979}
J.H.~Conway and S.P.~Norton, \emph{Monstrous {Moonshine}}, \href{https://doi.org/10.1112/blms/11.3.308}{\emph{Bulletin of the London Mathematical Society} {\bfseries 11} (1979) 308}.

\bibitem{thompson_numerology_1979}
J.G.~Thompson, \emph{Some {Numerology} between the {Fischer}-{Griess} {Monster} and the {Elliptic} {Modular} {Function}}, \href{https://doi.org/10.1112/blms/11.3.352}{\emph{Bulletin of the London Mathematical Society} {\bfseries 11} (1979) 352}.

\bibitem{frenkel_vertex_1989}
I.~Frenkel, J.~Lepowsky and A.~Meurman, \emph{Vertex operator algebras and the Monster}, Academic press (1989).

\bibitem{borcherds_monstrous_1992}
R.E.~Borcherds, \emph{Monstrous moonshine and monstrous {Lie} superalgebras}, \href{https://doi.org/10.1007/BF01232032}{\emph{Inventiones Mathematicae} {\bfseries 109} (1992) 405}.

\bibitem{lin_duality_2021}
Y.-H.~Lin and S.-H.~Shao, \emph{Duality defect of the monster {CFT}}, {\emph{Journal of Physics A: Mathematical and Theoretical} {\bfseries 54} (2021) 065201}.

\bibitem{bae_conformal_2021}
J.-B.~Bae, J.A.~Harvey, K.~Lee, S.~Lee and B.C.~Rayhaun, \emph{Conformal {Field} {Theories} with {Sporadic} {Group} {Symmetry}}, \href{https://doi.org/10.1007/s00220-021-04207-7}{\emph{Communications in Mathematical Physics} {\bfseries 388} (2021) 1}.

\bibitem{Cheng:2012tq}
M.C.N.~Cheng, J.F.R.~Duncan and J.A.~Harvey, \emph{{Umbral Moonshine}}, \href{https://doi.org/10.4310/CNTP.2014.v8.n2.a1}{\emph{Commun. Num. Theor. Phys.} {\bfseries 08} (2014) 101} [\href{https://arxiv.org/abs/1204.2779}{{\ttfamily 1204.2779}}].

\bibitem{Duncan:2021ith}
J.F.R.~Duncan, J.A.~Harvey and B.C.~Rayhaun, \emph{{An Overview of Penumbral Moonshine}},  \href{https://arxiv.org/abs/2109.09756}{{\ttfamily 2109.09756}}.

\bibitem{hohn_selbstduale_1995}
G.~H{\"o}hn, \emph{Selbstduale {Vertexoperatorsuperalgebren} und das {Babymonster} ({Self}-dual {Vertex} {Operator} {Super} {Algebras} and the {Baby} {Monster})}, Ph.D. thesis, Bonn University, 1995.

\bibitem{norton_anatomy_1998}
S.P.~Norton, \emph{Anatomy of the {Monster}: {I}},  in \emph{The {Atlas} of {Finite} {Groups} -- {Ten} {Years} {On}}, pp.~198--214, Cambridge University Press (1998).

\bibitem{harvey_hecke_2018}
J.A.~Harvey and Y.~Wu, \emph{Hecke {Relations} in {Rational} {Conformal} {Field} {Theory}}, \href{https://doi.org/10.1007/JHEP09(2018)032}{\emph{Journal of High Energy Physics} {\bfseries 2018} (2018) 32}.

\bibitem{hohn_mckays_2012}
G.~H{\"o}hn, C.H.~Lam and H.~Yamauchi, \emph{Mckay’s {$E_6$} {Observation} on the {Largest} {Fischer} {Group}}, {\emph{Communications in Mathematical Physics} {\bfseries 310} (2012) 329}.

\bibitem{haghighat_topological_2023}
B.~Haghighat and Y.~Sun, \emph{Topological {Defect} {Lines} in bosonized {Parafermionic} {CFTs}},  2023.

\bibitem{chang_topological_2019}
C.-M.~Chang, Y.-H.~Lin, S.-H.~Shao, Y.~Wang and X.~Yin, \emph{Topological defect lines and renormalization group flows in two dimensions}, {\emph{Journal of High Energy Physics} {\bfseries 2019} (2019) 1}.

\bibitem{miyamoto_vertex_2003}
M.~Miyamoto, \emph{Vertex operator algebras generated by two conformal vectors whose $\tau$-involutions generate {$S_3$}}, {\emph{Journal of Algebra} {\bfseries 268} (2003) 653}.

\bibitem{sakuma_vertex_2003}
S.~Sakuma and H.~Yamauchi, \emph{Vertex operator algebra with two {Miyamoto} involutions generating {$S_3$}}, {\emph{Journal of Algebra} {\bfseries 267} (2003) 272}.

\bibitem{lam_mckay_2005}
C.H.~Lam, H.~Yamada and H.~Yamauchi, \emph{Mckay's observation and vertex operator algebras generated by two conformal vectors of central charge 1/2}, {\emph{International Mathematics Research Papers} {\bfseries 2005} (2005) 117}.

\bibitem{verlinde_fusion_1988}
E.~Verlinde, \emph{Fusion rules and modular transformations in 2d conformal field theory}, {\emph{Nuclear Physics B} {\bfseries 300} (1988) 360}.

\bibitem{cummins_congruence_2003}
C.J.~Cummins and S.~Pauli, \emph{Congruence {Subgroups} of {$PSL(2, \mathbb{Z})$} of {Genus} {Less} than or {Equal} to 24}, {\emph{Experimental mathematics} {\bfseries 12} (2003) 243}.

\bibitem{sagemath}
{The Sage Developers}, \emph{{S}ageMath, the {S}age {M}athematics {S}oftware {S}ystem ({V}ersion 2.3.2)}, 2024.

\bibitem{paquette_monstrous_2016}
N.M.~Paquette, D.~Persson and R.~Volpato, \emph{Monstrous {BPS}-{Algebras} and the {Superstring} {Origin} of {Moonshine}}, {\emph{Communications in Number Theory and Physics} {\bfseries 10} (2016) 433}.

\bibitem{paquette_bps_2017}
N.M.~Paquette, D.~Persson and R.~Volpato, \emph{Bps algebras, genus zero and the heterotic {Monster}}, {\emph{Journal of Physics A: Mathematical and Theoretical} {\bfseries 50} (2017) 414001}.

\end{thebibliography}\endgroup

\end{document}
\typeout{get arXiv to do 4 passes: Label(s) may have changed. Rerun}